\documentclass[journal]{IEEEtran}
\IEEEoverridecommandlockouts
\usepackage{cite}
\usepackage{amsmath,amssymb,amsfonts}
\usepackage{graphicx}
\usepackage{textcomp}
\usepackage{xcolor}
\usepackage{balance}
\usepackage{cite}
\usepackage{hyperref}
\usepackage{amsmath,amssymb,amsfonts}
\usepackage{amsmath}
\usepackage{amsthm}

\DeclareMathOperator*{\argmin}{argmin}
\usepackage{graphicx}
\usepackage{textcomp}
\usepackage{xcolor}
\usepackage{graphicx}
\usepackage{float}
\usepackage{subfigure}
\usepackage{amsmath}
\usepackage{amsfonts,amssymb}
\usepackage{mathrsfs}
\usepackage{mathtools}
\usepackage{bm}
\usepackage{multirow}
\usepackage{array}
\usepackage{amssymb}
\usepackage{amsmath}
\usepackage{cite}
\usepackage{url}
\usepackage{xcolor}
\usepackage{cite,graphicx,amsmath,amssymb}
\usepackage{subfigure}
\usepackage{fancyhdr}
\usepackage{mdwmath}
\usepackage{mdwtab}
\usepackage{caption}
\usepackage{amsthm}
\usepackage{setspace}
\usepackage{bm}
\usepackage{mathtools}
\usepackage{dsfont}
\usepackage{bbm}
\usepackage{algorithm}
\usepackage{algorithmic}
\newtheorem{remark}{Remark}
\newtheorem{theorem}{Theorem}

\newtheorem{lemma}{Lemma}

\newtheorem{corollary}{Corollary}

\makeatletter
\newcommand{\biggg}{\bBigg@{3}}
\newcommand{\Biggg}{\bBigg@{3.5}}
\makeatother
\def\BibTeX{{\rm B\kern-.05em{\sc i\kern-.025em b}\kern-.08em
    T\kern-.1667em\lower.7ex\hbox{E}\kern-.125emX}}
\expandafter\def\expandafter\normalsize\expandafter{%
    \normalsize%
    \setlength\abovedisplayskip{4pt}%
    \setlength\belowdisplayskip{4pt}%
    \setlength\abovedisplayshortskip{2pt}%
    \setlength\belowdisplayshortskip{2pt}%
}
\begin{document}
\title{Secure Wireless Communications via Frequency Diverse Arrays}
\author{Zhenqiao Cheng, Chongjun Ouyang, and Xingqi Zhang\vspace{-10pt}
\thanks{Z. Cheng is with the 6G Research Centre, China Telecom Beijing Research Institute, Beijing 102209, China (e-mail: chengzq@chinatelecom.cn).}
\thanks{C. Ouyang was with University College Dublin, Dublin, Ireland, and is now with Queen Mary University of London, London, U.K. (e-mail: c.ouyang@qmul.ac.uk).}
\thanks{X. Zhang is with Department of Electrical and Computer Engineering, University of Alberta, Edmonton AB, T6G 2R3, Canada (email: xingqi.zhang@ualberta.ca).}}
\maketitle
\begin{abstract}
A novel frequency diverse array (FDA)-assisted secure transmission framework is proposed, which leverages additional frequency offsets to enhance physical layer security. Specifically, an FDA-assisted wiretap channel is considered, where the transmit beamforming and frequency offsets at each antenna are jointly optimized. A novel alternating optimization-based method is introduced to address the non-convex problem of secure transmission, focusing on minimizing transmit power and maximizing the secrecy rate. Numerical results are provided to demonstrate the superiority of the FDA-based framework compared to systems employing traditional phased array antennas in secure transmission.
\end{abstract}
\section{Introduction}
Physical layer security (PLS) is an attractive paradigm for achieving secure communications \cite{b17,Khisti2010}. In the field of PLS, the frequency diverse array (FDA) represents a key hardware architecture for implementing practical PLS-oriented beamforming \cite{b3}. By appropriately programming the frequency offset of each array element in an FDA, a joint range-and-direction beampattern can be achieved, in contrast to the direction-only beampattern realized by conventional phased arrays \cite{b3}. As a result, FDA beamforming enables secure, high-data-rate transmission across both the range and direction dimensions, whereas phased array systems can only secure communications in the direction dimension.

Building on this background, FDA-based PLS design has been studied in the current literature. The work in \cite{b18} investigated FDA-based directional modulation (DM) using random frequency offsets along with artificial noise. This approach was later extended to other DM scenarios; see \cite{b16,Cheng2021,Jian2023} and relevant references. In addition to DM, there are also studies exploring FDA-based beamforming aimed at improving the secrecy transmission rate. For instance, the authors in \cite{b2} designed beamforming techniques for FDAs to improve the secrecy rate. The work in \cite{Ouyang2020} further analyzed the average secrecy rate achieved by FDAs using linearly varying frequency offsets. The work in \cite{Cheng2024} investigated the use of FDA in covert communications to enhance PLS. More recently, FDA has been integrated with emerging movable antenna technologies to further strengthen wireless PLS \cite{Cheng2025,Cheng2026}.

In contrast to these initial results, this article provides a thorough investigation of the secrecy performance achieved by FDAs. The contributions are listed as follows: {\romannumeral1}) We propose an FDA-based secure transmission framework that jointly optimizes frequency offsets and transmit beamforming to enhance PLS. {\romannumeral2}) We address the problem of transmit power minimization while ensuring a target secrecy rate, which has not been explored in existing works, and propose an alternating optimization-based method to solve this non-convex problem. {\romannumeral3}) We also consider the maximization of the secrecy rate under a power budget constraint. Although this problem has been discussed in \cite{b2}, our work adopts a more general channel model that does not require constant channel amplitudes. {\romannumeral4}) Simulation results demonstrate that the proposed FDA-based secure transmission framework offers a better secrecy performance than conventional phased arrays and FDAs without properly designed frequency offsets.

\begin{figure}[!t]
\centering
\setlength{\abovecaptionskip}{0pt}
\includegraphics[height=0.13\textwidth]{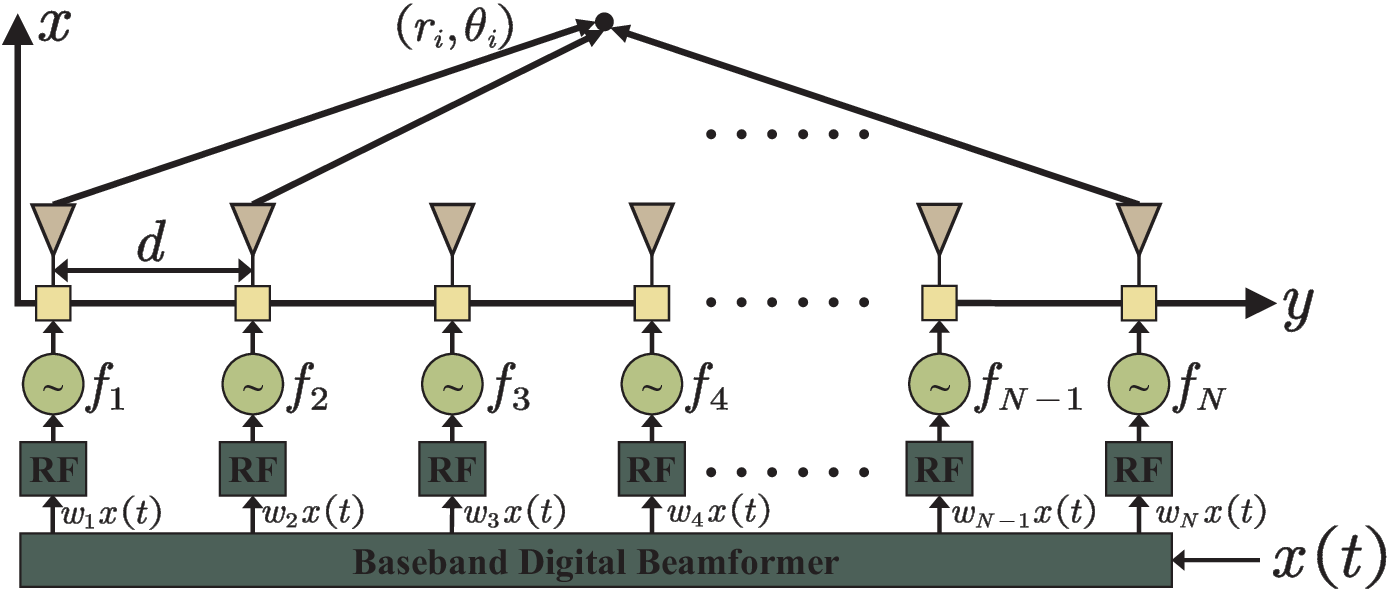}
\caption{The structure of an FDA, where $w_n$ is the digital beamforming coefficient and $f_{n}=f_{c}+\Delta{f_n}$ for $n=1,\cdots,N$.}
\label{figure_FDA}
\vspace{-15pt}
\end{figure}

\section{System Model}
In a multiple-antenna wiretap channel, Alice sends a secret message to Bob ($\rm{b}$), which is overheard by Eve ($\rm{e}$). Assume that Alice is equipped with an FDA consisting of $N$ antennas, whose structure is illustrated in {\figurename} \ref{figure_FDA}. Each receiver, Bob and Eve, is equipped with a single-antenna device. Let $(r_i\cos{\theta_i},r_i\sin{\theta_i})$ denote the two-dimensional coordinates of node $i\in\{{\rm{b}},{\rm{e}}\}$, where $r_i$ represents the propagation distance from the origin and $\theta_i$ denotes the associated directional angle. Let $d$ represent the inter-element spacing between antennas, and let $(x_0,0)$ denote the coordinates of the first antenna, as depicted in {\figurename} \ref{figure_FDA}. Thus, the channel response from the $n$th antenna to node $i$ at time instant $t$ is given as follows:
\begin{equation}\label{Transmit_Steering}
\begin{split}
	h_{i,n}(t,f_n)
=\frac{\lambda}{4\pi r_{i,n}}{\rm{e}}^{{\rm{j}}{2\pi}f_n(t-r_{i,n}/c)},\quad i\in\{{\rm{b}},{\rm{e}}\},
\end{split}
\end{equation}
where $f_n=f_c+\Delta f_n$ is the frequency used at the $n$th antenna element, with $f_c$ and $\Delta f_n$ being the carrier frequency and frequency offset, respectively. Moreover, $c$ is the speed of light, $\lambda=2\pi f_c$ is the wavelength of the carrier frequency, and
\begin{equation}
r_{i,n}=\sqrt{(x_0+(n-1)d-r_i\cos{\theta_i})^2+r_i^2\sin^2{\theta_i}}
\end{equation}
denotes the distance between node $i$ and the $n$th antenna element for $i\in\{{\rm{b}},{\rm{e}}\}$ and $n\in{\mathcal{N}}\triangleq\{1,\ldots,N\}$. 
\vspace{-5pt}
\begin{remark}
By setting $r_{i,n}=r_{i}$ for $n=1,\ldots,N$, the channel model in \eqref{Transmit_Steering} reduces to the one described in \cite{b2}. It is worth emphasizing that our considered channel model is more general than that in \cite{b2}, as it accounts for variations in channel power across the entire array. Therefore, the method proposed in \cite{b2} is not directly applicable to our model.
\end{remark}
\vspace{-5pt}
As a result, the received signals at Bob and Eve at time instant $t$ can be written as follows:
\begin{align}
  y_{i}\left(t\right)={\mathbf{h}}_i^{\mathsf{H}}(t,{\mathbf{f}}){\mathbf w}(t)x(t)+n_i(t),\quad i\in\{{\rm{b}},{\rm{e}}\},
\end{align}
where $x\left(t\right)\in{\mathbbmss C}$ denotes the transmitted symbol, satisfying ${\mathbbmss E}\{|x(t)|^2\}=1$, and ${\mathbf{h}}_i(t,{\mathbf{f}})=[h_{i,n}(t,f_n)]_{n=1}^{N}\in{\mathbbmss C}^{N\times1}$ represents the channel from the FDA to the receiver. The frequency vector is ${\mathbf{f}}=[f_1;\ldots;f_N]\in{\mathbbmss C}^{N\times1}$, ${\mathbf w}(t)\in{\mathbbmss C}^{N\times1}$ is the digital beamforming vector, and $n_i(t)\in{\mathcal{CN}}(0,\sigma_i^2)$ denotes the Gaussian noise with variance $\sigma_i^2$. The instantaneous received signal-to-noise ratio (SNR) at node $i$ is given by
\begin{align}
	\gamma_i\left(t\right)=\lvert{\mathbf{h}}_i^{\mathsf{H}}(t,{\mathbf{f}}){\mathbf w}(t)\rvert^2/{\sigma_i^2},\quad i\in\{{\rm{b}},{\rm{e}}\},
\end{align}
and the secrecy rate at time $t$ can be written as follows:
\begin{align}\label{Secrecy_Rate_General}
	{\mathcal R}_{\rm s}(t)=\max\{\log_2(1+\gamma_{\rm{b}}(t))-\log_2(1+\gamma_{\rm{e}}(t)),0\}.
\end{align}

Assume that Eve is a system registered user, and thus Alice has access to Eve's channel state information (CSI). Moreover, to explore the system's performance bounds, we assume that all CSI is perfectly known to Alice. Under this setup, we will discuss two basic scenarios of secure transmission.
\subsubsection{Transmit Power Minimization}
The transmit beamformer ${\mathbf w}(t)$ and the FDA beamforming frequencies $\mathbf{f}$ can be designed to minimize the system's transmit power, subject to the secrecy rate constraints. This can be formulated as follows:
\begin{align}\label{Problem1}
\min_{{\mathbf w}\left(t\right),{\mathbf{f}}}\|\mathbf w(t)\|^2~{\rm{s.t.}}~{\mathcal R}_{\rm s}(t)\geq R, \Delta f_n\in\left[0,f_{\mathsf{m}}\right],n\in{\mathcal{N}},\tag{$\mathcal{P}_1$}
\end{align}
where ${ R}>0$ is the target secrecy rate and $f_{\mathsf{m}}>0$ represents the maximum frequency offset.
\subsubsection{Secrecy Rate Maximization}
Besides, ${\mathbf w}(t)$ and $\mathbf{f}$ can be designed to maximize the secrecy rate, subject to the transmit power constraints. This can be formulated as follows:
\begin{align}\label{Problem2}
\max_{{\mathbf w}\left(t\right),{\mathbf f}}{\mathcal R}_{\rm s}(t)~{\rm{s.t.}}~\lVert{\mathbf w}(t)\rVert^2\leq P,\Delta f_n\in\left[0,f_{\mathsf{m}}\right],n\in{\mathcal{N}},\tag{${\mathcal P}_2$}
\end{align}
where $P>0$ is the power budget.
\section{Transmit Power Minimization}
\subsection{Transmit Beamforming Design}
Due to the tight coupling of ${\mathbf w}\left(t\right)$ and ${\mathbf f}$, we first consider the design of ${\mathbf w}(t)$ for a given ${\mathbf f}$. For clarity, we drop the time index $t$ when there is no ambiguity and denote $\hat{\mathbf{h}}_i={\mathbf{h}}_i({\mathbf{f}})\sigma_i^{-1}$ and ${\mathbf u}=p^{-\frac{1}{2}}{\mathbf w}$ with $p=\|{\mathbf w}\|^2$. Then the subproblem of optimizing ${\mathbf w}$ can be equivalently rewritten as follows:
\begin{align}\label{TPM_Sub1}
\min_{{\mathbf u},p}~p\quad{\rm{s.t.}}~\frac{{\mathbf u}^{\mathsf{H}}((1-2^R){\mathbf I}+p{\bm\Sigma}){\mathbf u}}
{{\mathbf u}^{\mathsf{H}}({\mathbf I}+p\hat{\mathbf{h}}_{\rm{e}}\hat{\mathbf{h}}_{\rm{e}}^{\mathsf{H}}){\mathbf u}}\geq0,\|{\mathbf u}\|^2=1,\tag{${\mathcal{P}}_{{\mathbf{w}}}^1$}
\end{align}
where ${\bm\Sigma}=\hat{\mathbf{h}}_{\rm{b}}\hat{\mathbf{h}}_{\rm{b}}^{\mathsf{H}}-2^R\hat{\mathbf{h}}_{\rm{e}}\hat{\mathbf{h}}_{\rm{e}}^{\mathsf{H}}\in{\mathbbmss{C}}^{N\times N}$.
\subsubsection{Optimal Solution to \eqref{TPM_Sub1}}
Since $p\geq0$, ${\mathbf I}+p\hat{\mathbf{h}}_{\rm{e}}\hat{\mathbf{h}}_{\rm{e}}^{\mathsf{H}}\succeq{\mathbf{0}}$, which yields ${\mathbf u}^{\mathsf{H}}({\mathbf I}+p\hat{\mathbf{h}}_{\rm{e}}\hat{\mathbf{h}}_{\rm{e}}^{\mathsf{H}}){\mathbf u}\geq0$, and equality holds only when ${\mathbf u}={\mathbf 0}$. Consequently, the first constraint of \eqref{TPM_Sub1} is equivalent to
${\mathbf u}^{\mathsf{H}}((1-2^R){\mathbf I}+p{\bm\Sigma}){\mathbf u}\geq 0$. Let ${\mathbf U}{\bm\Lambda}{\mathbf U}^{\mathsf{H}}$ denote the eigen-decomposition (EVD) of ${\bm\Sigma}$, where ${\mathbf U}{\mathbf U}^{\mathsf{H}}={\mathbf I}$ and ${\bm\Lambda}={\mathrm{diag}}\left\{\sigma_1,\cdots,\sigma_N\right\}$, with $\sigma_1\geq\cdots\geq\sigma_N$ representing the eigenvalues of $\bm\Sigma$. Since $\|{\mathbf u}\|^2=1$, we have
\begin{equation}
\begin{split}
{\mathbf u}^{\mathsf{H}}((1-2^R){\mathbf I}+p{\bm\Sigma}){\mathbf u}\in[1-2^R+p\sigma_1,1-2^R+p\sigma_N]\nonumber.
\end{split}
\end{equation}
When \eqref{TPM_Sub1} is feasible, $1-2^R+p\sigma_1\geq0$ must be satisfied, which yields $\sigma_1\geq\frac{2^R-1}{p}>0$, and thus $p\geq\frac{2^R-1}{\sigma_1}>0$. Taken together, the minimum value of $p$ is given by $\frac{2^R-1}{\sigma_1}$, and the optimal ${\mathbf u}$ is the normalized principal eigenvector of ${\bm\Sigma}$. This also implies that minimizing $p$ is equivalent to maximizing $\lambda_1$.
\subsubsection{Calculation of the Principal Eigenvalue}
Using the matrix determinant lemma, a closed-form solution for the principal eigenvalue $\lambda_1$ can be derived. Due to page limitations, we omit the detailed steps. Specifically, we have $\lambda_{1} = {-\frac{1}{2}w_1+\frac{1}{2}\sqrt{w_1^2+2^{2+R}w_2}}$, where $w_1=2^R \lVert\hat{\mathbf{h}}_{\rm{e}}\rVert^2-\lVert\hat{\mathbf{h}}_{\rm{b}}\rVert^2$ and $w_2=\lVert\hat{\mathbf{h}}_{\rm{b}}\rVert^2\lVert\hat{\mathbf{h}}_{\rm{e}}\rVert^2-\lvert\hat{\mathbf{h}}_{\rm{e}}^{\mathsf{H}}\hat{\mathbf{h}}_{\rm{b}}\rvert^2$. Recalling \eqref{Transmit_Steering}, we have 
\begin{align}
\lVert\hat{\mathbf{h}}_{i}\rVert^2=\sum\nolimits_{n=1}^{N}\frac{\lvert h_{i,n}(t,f_n)\rvert^2}{\sigma_i^2}=\sum\nolimits_{n=1}^{N}\frac{\lambda^2}{(4\pi r_{i,n})^2\sigma_i^2}.
\end{align}
This indicates that $\lVert\hat{\mathbf{h}}_{i}\rVert^2$ is not influenced by $\mathbf{f}$ or $t$. Therefore, maximizing $\lambda_1$ is equivalent to minimizing $\lvert\hat{\mathbf{h}}_{\rm{e}}^{\mathsf{H}}\hat{\mathbf{h}}_{\rm{b}}\rvert^2$. When $\lvert\hat{\mathbf{h}}_{\rm{e}}^{\mathsf{H}}\hat{\mathbf{h}}_{\rm{b}}\rvert^2=0$, we obtain an upper bound for $\lambda_1$ as follows:
\begin{align}
\lambda_1\leq -\frac{1}{2}w_1+\frac{1}{2}(2^R \lVert\hat{\mathbf{h}}_{\rm{e}}\rVert^2+\lVert\hat{\mathbf{h}}_{\rm{b}}\rVert^2)=\lVert\hat{\mathbf{h}}_{\rm{b}}\rVert^2. 
\end{align}
This provides the lower bound for the transmit power, which is $\lVert{\mathbf{w}}\rVert^2=({2^R-1})/{\lVert\hat{\mathbf{h}}_{\rm{b}}\rVert^2}$.
\subsubsection{Equivalent Transformation of \eqref{Problem1}}
The above arguments imply that minimizing the transmit power is equivalent to minimizing $\lvert\hat{\mathbf{h}}_{\rm{e}}^{\mathsf{H}}\hat{\mathbf{h}}_{\rm{b}}\rvert^2$, or equivalently, $g(t,{\mathbf f})=\lvert{\mathbf{h}}_{\rm{e}}^{\mathsf{H}}(t,{\mathbf f}){\mathbf{h}}_{\rm{b}}(t,{\mathbf f})\rvert^2$. This can be formulated as follows:
\begin{align}\label{Problem4}
\min\nolimits_{{\mathbf f}}~g(t,{\mathbf f})\quad{\rm{s.t.}}~\Delta f_n\in\left[0,f_{\mathsf{m}}\right],n\in{\mathcal{N}}.\tag{${\mathcal{P}}_{\mathbf{f}}$}
\end{align}
After obtaining the optimized $\mathbf{f}$, the digital beamformer $\mathbf{w}(t,{\mathbf f})$ can be set as the principal eigenvalue of 
\begin{align}
\hat{\mathbf{h}}_{\rm{b}}(t,{\mathbf f})\hat{\mathbf{h}}_{\rm{b}}^{\mathsf{H}}(t,{\mathbf f})-2^R\hat{\mathbf{h}}_{\rm{e}}(t,{\mathbf f})\hat{\mathbf{h}}_{\rm{e}}^{\mathsf{H}}(t,{\mathbf f})\triangleq{\bm\Sigma}(t,{\mathbf f}).
\end{align} 
\subsection{FDA Beamforming Design}
We next aim to solve problem \eqref{Problem4}. Based on \eqref{Transmit_Steering}, we have
\begin{equation}\label{Objective_Func1}
\begin{split}
&g(t,{\mathbf f})
=\frac{\lambda^2}{(4\pi)^2}\left|\sum\nolimits_{n=1}^{N}\alpha_n{\rm e}^{{\rm{j}}\omega_nf_n}\right|^2=\frac{\lambda^2}{(4\pi)^2}\sum_{n=1}^{N}\alpha_n^2\\
&+\frac{\lambda^2}{(4\pi)^2}\sum_{n=1}^{N}\sum_{n'\neq n}\alpha_n\alpha_{n'}\cos(\omega_nf_n-\omega_{n'}f_{n'})\triangleq g({\mathbf f}),
\end{split}
\end{equation}
where $w_n=\frac{2\pi(r_{{\rm{e}},n}-r_{{\rm{b}},n})}{c}$ and $\alpha_n=\frac{1}{r_{{\rm{b}},n}r_{{\rm{e}},n}}$. The results in \eqref{Objective_Func1} suggest that the optimized $\mathbf{f}$ is time-independent, and thus the corresponding transmit power is also time-independent.

Problem \eqref{Problem4} is NP-hard, making the optimal solution challenging to find. As a compromise, we adopt an alternating optimization-based method to find a suboptimal solution, where each frequency value is treated as one block, and they are optimized alternately. Given the frequency values $\left\{f_{n'}\right\}_{n'\neq k}$, the resultant optimization problem is given by
\begin{align}\label{Problem6}
	\min\nolimits_{{f}_n}~g_n(f_n)\quad{\rm{s.t.}}~f_n\in\left[f_c,f_c+f_{\mathsf{m}}\right],\tag{${\mathcal{P}}_{f_n}$} 
\end{align}
where $g_n(f_n)=\sum_{n'\neq n}\alpha_{n'}\cos(\omega_nf_n-\omega_{n'}f_{n'})$. Note that $g_n(f_n)$ can also be written as follows: 
\begin{align}
g_n(f_n)=\sqrt{A_n^2+B_n^2}\cos(|w_n|f_n-\varphi_n),
\end{align}
where $A_n=\sum_{n'\neq n}\alpha_{n'}\cos(\omega_{n'}f_{n'})$, and $B_n=\sum_{n'\neq n}\alpha_{n'}\sin(\omega_{n'}f_{n'})$, and $\varphi_n={\rm{sgn}}(w_n){\rm{atan2}}({{B}_n},{A}_n)$. Problem \eqref{Problem6} is thus equivalent to
\begin{align}\label{Optimal_Frequency_Offset}
f_n^{\star}=\argmin\nolimits_{({|w_n|f_c-\varphi_n})\leq x\leq({|w_n|(f_c+f_{\mathsf{m}})-\varphi_n})}\cos(x).
\end{align}
The optimal solution is given as follows:
\begin{align}
f_n^{\star}=
\begin{cases}
{f_c+f_{\mathsf{m}}}\!\!& {0\leq a_n\leq\pi,c_n+a_n<\pi}\\
\frac{\pi+d_n}{|\omega_n|}\!\!& {0\leq a_n\leq\pi,c_n+a_n\geq\pi}\\
{f_c}\!\!& {\pi\leq a_n<2\pi,c_n+2a_n<4\pi}\\
{f_c+f_{\mathsf{m}}}\!\!& {\pi\leq a_n<2\pi,4\pi-a_n<c_n+a_n<3\pi}\\
\frac{3\pi+d_n}{|\omega_n|}\!\!& {\pi\leq a_n<2\pi,c_n+a_n\geq3\pi}
\end{cases},\nonumber
\end{align}
where $a_n=b_n \bmod 2\pi$ with $b_n=|\omega_n|f_c-\varphi_n$, $c_n=|\omega_n|f_{\mathsf{m}}$, and $d_n=b_n-a_n+\varphi_n$.

\begin{algorithm}[!t]
\algsetup{linenosize=\tiny} \scriptsize
  \caption{The proposed method for solving \eqref{Problem4}}
  \label{Algorithm1}
  \begin{algorithmic}[1]
    \STATE Initialize ${\mathbf f}={\mathbf f}^{(0)}$ and set $k=0$;
    \REPEAT
      \FORALL{$n=1$ to $N$}
      \STATE Update the frequency offset $f_n^{(k+1)}$ by \eqref{Optimal_Frequency_Offset};
      \ENDFOR
      \STATE Update the iteration index $k\leftarrow{k+1}$;
    \UNTIL{convergence}.
  \end{algorithmic}
\end{algorithm}

The proposed method is summarized in Algorithm \ref{Algorithm1}. Since each block of the alternating optimization uses the closed-form globally optimal solutions for \eqref{Problem6}, the objective function decreases monotonically. Furthermore, as the objective function is lower-bounded by zero, these two properties together guarantee the convergence of the proposed method. It is worth noting that although $f_n^{\star}$ yields the conditionally optimal frequency offsets when the other offsets are fixed, the final algorithm only achieves a suboptimal solution. Finding a low-complexity method to obtain the globally optimal frequency offsets for all antennas remains an open problem. Nevertheless, as demonstrated in the simulation results, the proposed method can approach the performance bound of the secrecy rate in certain scenarios---particularly when $N$ is large---indicating that the proposed approach achieves a near-optimal solution.
\subsection{Average Transmit Power Minimization}
In practice, it is also important to address the problem of average power minimization, defined as follows:
\begin{subequations}\label{ATPM}
\begin{align}
&\min\nolimits_{\left\{{\mathbf w}(t)\right\}_{t\in[0,T]},{\mathbf{f}}}~\frac{1}{T}\int_{0}^{T}\|\mathbf w(t)\|^2{\rm d}t\\
    &~{\rm{s.t.}}~{\mathcal R}_{\rm s}(t)\geq R,\forall t\in[0,T],f_n\in\left[0,f_{\mathsf{m}}\right],\forall n.
    \end{align}
\end{subequations}
This problem minimizes the average transmit power over a predefined time period of length $T$. For brevity, assume that $T$ is shorter than the channel coherence time. 

As discussed previously, the solution to \eqref{Problem4} is time-invariant, given the channel responses. Thus, \eqref{ATPM} has the same solution as \eqref{Problem1}, which can be achieved by first calculating $\mathbf f$ from \eqref{Problem4}, and then updating ${\mathbf w}(t)$ as the principal eigenvector of ${\bm\Sigma}(t,{\mathbf f})$. Let ${\mathbf f}^{\star}$ denote the optimized frequency vector from Algorithm \ref{Algorithm1}. The average transmit power is then given by
\begin{align}\label{Optimal_TPM_Solution}
\overline{P}^{\star}=\frac{1}{T}\int_{0}^{T}\frac{2^R-1}{\lambda_{{\bm\Sigma}\left(t,{\mathbf f}^{\star}\right)}}{\rm d}t=\frac{2^R-1}{\lambda_{{\bm\Sigma}\left(t,{\mathbf f}^{\star}\right)}}\geq
\frac{2^R-1}{\lVert\hat{\mathbf{h}}_{\rm{b}}\rVert^2},
\end{align}
where the second equality holds because $\lambda_{{\bm\Sigma}\left(t,{\mathbf f}^{\star}\right)}$, the principal eigenvalue of ${\bm\Sigma}\left(t,{\mathbf f}^{\star}\right)$, is time-independent. 

However, from the perspective of practical implementation, updating ${\mathbf w}\left(t\right)$ in real time is challenging, as it lacks a closed-form solution, requiring an EVD. As a compromise, we consider a maximal ratio transmission (MRT)-based scheme, where ${\mathbf w}(t)=\frac{\sqrt{P}{\mathbf h}_{\rm b}(t,{\mathbf f})}{\|{\mathbf h}_{\rm b}(t,{\mathbf f})\|}$, which provides a closed-form solution, making it easier to implement compared to the EVD-based scheme\footnote{Real-time optimization of frequency offsets and beamformers presents significant challenges in practical FDA systems. The proposed average design aims to mitigate this issue by deriving time-independent frequency offsets and closed-form beamformers, thereby facilitating real-time implementation. However, to achieve a more robust design, we shall consider the time variation of the wireless channel. While important, this is beyond the scope of this work. Interested readers are referred to \cite{b2} and \cite{Cheng2025} for further insights.}. The secrecy rate for the MRT scheme can be written as ${\mathcal R}_{\mathsf{mrt}}=\log_2{\left(\frac{1+P\|{\mathbf h}_{\rm b}(t,{\mathbf f})\|^2}
{1+Pg({\mathbf f})/\|{\mathbf h}_{\rm b}(t,{\mathbf f})\|^2}\right)}$. Note that ${\mathcal R}_{\mathsf{mrt}}$ increases monotonically with $P$, and thus we require $P\geq\frac{2^{ R}-1}{\|{\mathbf h}_{\rm b}(t,{\mathbf f})\|^2-2^{R}g\left({\mathbf f}\right)/\|{\mathbf h}_{\rm b}(t,{\mathbf f})\|^2}$ to let ${\mathcal R}_{\mathsf{mrt}}\geq{R}$. Consequently, minimizing the transmit power is equivalent to minimizing $g\left({\mathbf f}\right)$, which corresponds to solving problem \eqref{Problem4}. The resulting optimal frequency offsets are thus identical to those obtained from the instantaneous transmit power minimization problem. Accordingly, the average transmit power is expressed as $P_{\mathsf{mrt}}=\frac{2^{ R}-1}{\|{\mathbf h}_{\rm b}(t,{\mathbf f}^{\star})\|^2-2^{R}g\left({\mathbf f}^{\star}\right)/\|{\mathbf h}_{\rm b}(t,{\mathbf f}^{\star})\|^2}$. Since $\overline{P}^{\star}$ in \eqref{Optimal_TPM_Solution} is obtained via the optimal transmit beamformer, we have $\overline{P}^{\star}\leq P_{\mathsf{mrt}}$. When $g\left({\mathbf f}^{\star}\right)=0$, i.e., Bob’s and Eve’s channels are orthogonal, both the EVD-based and MRT-based schemes yield the performance lower bound of problem \eqref{Problem1}.

\section{Secrecy Rate Maximization}
\subsection{Transmit Beamforming Design}
When problem \eqref{Problem2} is feasible, it can be observed from \eqref{Secrecy_Rate_General} that the secrecy rate increases monotonically with $\lVert{\mathbf w}(t)\rVert^2$. Hence, when the secrecy rate is maximized, it holds that $\lVert{\mathbf w}(t)\rVert^2=P$. On this basis, the marginal problem with respect to ${\mathbf w}(t)$, given $\mathbf{f}$, can be formulated as follows:
\begin{align}\label{STRM_Sub1}
\max_{{\mathbf w}}~\frac{{\mathbf{w}}^{\mathsf{H}}(P^{-1}{\mathbf{I}}+\hat{\mathbf{h}}_{\rm{b}}\hat{\mathbf{h}}_{\rm{b}}^{\mathsf{H}}){\mathbf{w}}}
{{\mathbf{w}}^{\mathsf{H}}(P^{-1}{\mathbf{I}}+\hat{\mathbf{h}}_{\rm{e}}\hat{\mathbf{h}}_{\rm{e}}^{\mathsf{H}}){\mathbf{w}}}\quad{\rm{s.t.}}~\lVert{\mathbf w}\rVert^2=P. \tag{${\mathcal{P}}_{{\mathbf{w}}}^2$}
\end{align}
For brevity, the notation $(t,{\mathbf{f}})$ is omitted. Problem \eqref{STRM_Sub1} is a Rayleigh quotient, and its solution satisfies
\begin{align}\label{SR_Optimal_Vector}
{\mathbf w}^{\star}=\sqrt{P}\frac{(P^{-1}{\mathbf{I}}+\hat{\mathbf{h}}_{\rm{e}}\hat{\mathbf{h}}_{\rm{e}}^{\mathsf{H}})^{-\frac{1}{2}}{\mathbf p}_{\bm\Delta}}{{\mathbf p}_{\bm\Delta}^{\mathsf{H}}(P^{-1}{\mathbf{I}}+\hat{\mathbf{h}}_{\rm{e}}\hat{\mathbf{h}}_{\rm{e}}^{\mathsf{H}})^{-1}{\mathbf p}_{\bm\Delta}},
\end{align}
where ${\mathbf p}_{\bm\Delta}\in{\mathbbmss{C}}^{N\times1}$ is the principal eigenvector of the matrix ${\bm\Delta}=(\frac{1}{P}{\mathbf{I}}+\hat{\mathbf{h}}_{\rm{e}}\hat{\mathbf{h}}_{\rm{e}}^{\mathsf{H}})^{-\frac{1}{2}}(\frac{1}{P}{\mathbf{I}}+\hat{\mathbf{h}}_{\rm{b}}\hat{\mathbf{h}}_{\rm{b}}^{\mathsf{H}})
(\frac{1}{P}{\mathbf{I}}+\hat{\mathbf{h}}_{\rm{e}}\hat{\mathbf{h}}_{\rm{e}}^{\mathsf{H}})^{-\frac{1}{2}}_{\bm\Delta}\in{\mathbbmss{C}}^{N\times N}$. Additionally, the maximum of the objective function in \eqref{STRM_Sub1} equals the principal eigenvalue of ${\bm\Delta}$.

The next task is to calculate the eigenvalues of ${\bm\Delta}$, which can be handled using the matrix determinant lemma. Due to space limitations, we report only the final result. Specifically, the principal eigenvalue of ${\bm\Delta}$ is given by $\lambda_{\bm\Delta}=1+\frac{P}{2}\frac{f_1+\sqrt{{f_1^2}+f_2}}{1+P\lVert\hat{\mathbf{h}}_{\rm{e}}\rVert^2}$, where $f_1=P(\lVert\hat{\mathbf{h}}_{\rm{b}}\rVert^2\lVert\hat{\mathbf{h}}_{\rm{e}}\rVert^2-\lvert\hat{\mathbf{h}}_{\rm{e}}^{\mathsf{H}}\hat{\mathbf{h}}_{\rm{b}}\rvert^2)+\lVert\hat{\mathbf{h}}_{\rm{b}}\rVert^2-\lVert\hat{\mathbf{h}}_{\rm{e}}\rVert^2$ and $f_2=4(1+P\lVert\hat{\mathbf{h}}_{\rm{e}}\rVert^2)(\lVert\hat{\mathbf{h}}_{\rm{b}}\rVert^2\lVert\hat{\mathbf{h}}_{\rm{e}}\rVert^2-\lvert\hat{\mathbf{h}}_{\rm{e}}^{\mathsf{H}}\hat{\mathbf{h}}_{\rm{b}}\rvert^2)$. As stated before, $\lVert\hat{\mathbf{h}}_{\rm{b}}\rVert^2$ and $\lVert\hat{\mathbf{h}}_{\rm{e}}\rVert^2$ are independent of $t$ and $\mathbf f$. Let $x=\lVert\hat{\mathbf{h}}_{\rm{b}}\rVert^2\lVert\hat{\mathbf{h}}_{\rm{e}}\rVert^2-\lvert\hat{\mathbf{h}}_{\rm{e}}^{\mathsf{H}}\hat{\mathbf{h}}_{\rm{b}}\rvert^2\geq0$. Then, we have $\frac{{\rm{d}}\lambda_{\bm\Delta}}{{\rm{d}}x}=\frac{P{(f_1+\sqrt{f_1^2+f_2})}+{2(1+P\lVert\hat{\mathbf{h}}_{\rm{e}}\rVert^2)}}
{2(1+P\lVert\hat{\mathbf{h}}_{\rm{e}}\rVert^2)\sqrt{f_1^2+f_2}/P}$, which, together with the fact that $f_2\geq0$, yields $\frac{{\rm{d}}\lambda_{\bm\Delta}}{{\rm{d}}x}\geq0$. Therefore, maximizing $\lambda_{\bm\Delta}$ is equivalent to minimizing $g({\mathbf{f}})=\lvert\hat{\mathbf{h}}_{\rm{e}}^{\mathsf{H}}\hat{\mathbf{h}}_{\rm{b}}\rvert^2$, which corresponds to the problem \eqref{Problem4} and can be efficiently solved using Algorithm \ref{Algorithm1}. Hence, we conclude that, under our considered setup, the optimal frequency offsets that maximize the secrecy rate are identical to those that minimize the transmit power. Let ${\textbf f}^{\star}$ denote the optimized frequency vector. Then, the secrecy rate can be expressed as $\log_2{\lambda_{\bm\Delta}}$, which is time-independent.
\subsection{Average Secrecy Rate Maximization}
The problem of average secrecy rate maximization can be formulated as follows:
\begin{subequations}\label{ASRM}
\begin{align}
&\max\nolimits_{\left\{{\mathbf w}(t)\right\}_{t\in\left[0,T\right]},{\mathbf{f}}}~\frac{1}{T}\int_{0}^{T}{\mathcal R}_{\rm s}(t){\rm d}t\label{ASTRM}\\
    &~{\rm{s.t.}}~\lVert{\mathbf w}(t)\rVert^2\leq P,\forall t\in[0,T],f_n\in\left[0,f_{\mathsf{m}}\right],\forall n,
\end{align}
\end{subequations}
which maximizes the average secrecy rate over a time period of length $T$. Since the instantaneous secrecy rate maximization problem \eqref{Problem2} has time-invariant optimized frequency offsets and secrecy rate, problem \eqref{ASRM} has the same optimal objective value as \eqref{Problem2}. This can be achieved by first designing $\mathbf f$ via Algorithm \ref{Algorithm1} and then updating ${\mathbf w}\left(t\right)$ as in \eqref{SR_Optimal_Vector}. To avoid real-time EVD operations, we also consider an MRT-based scheme, where the resulting secrecy rate is given by $\log_2{\left(\frac{1+P\|{\mathbf h}_{\rm b}(t,{\mathbf f}^{\star})\|^2}
{1+Pg({\mathbf f}^{\star})/\|{\mathbf h}_{\rm b}(t,{\mathbf f}^{\star})\|^2}\right)}$.

\section{Numerical Results}
This section employs computer simulations to demonstrate the performance of the proposed beamforming schemes. Unless otherwise specified, the parameters are set as follows: $f_{0}=2.4$ GHz, $\sigma_{\rm b}^2=\sigma_{\rm e}^2=-100$ dBmW, $x_0=0$ m, $r_{\rm{b}}=r_{\rm{e}}-20~{\text{m}}\in[50~{\text{m}},150~{\text{m}}]$, ${\theta_{\rm b}}={\theta_{\rm e}}\in[0,\pi]$, $t=0$, and $f_{\mathsf{m}}=3$ MHz for $n\in{\mathcal{N}}$. The time period is set to $T=20~{\mu}{\text{s}}$, following the setup adopted in \cite{b2}. All results are obtained by averaging over $1000$ independent channel realizations.

\begin{figure}[!t]
\centering
\setlength{\abovecaptionskip}{0pt}
\includegraphics[height=0.175\textwidth]{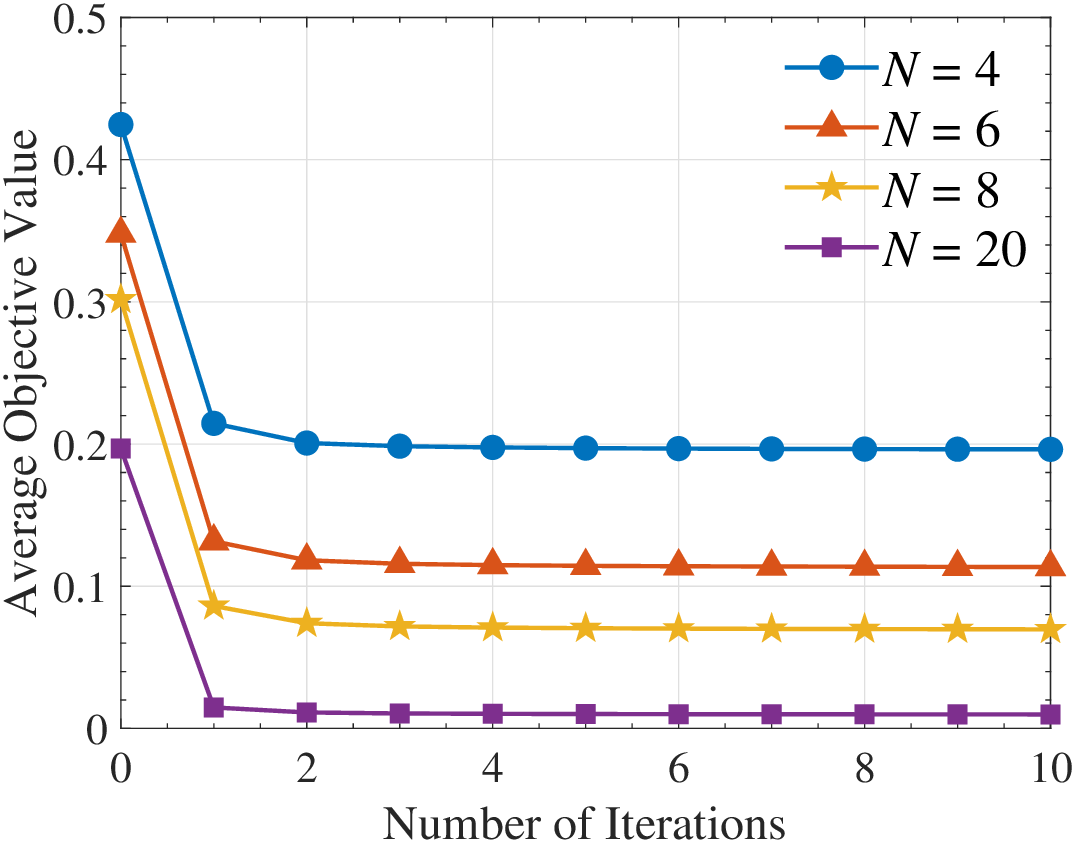}
\caption{Average convergence performance of Algorithm \ref{Algorithm1}.}
\label{figure1}
\vspace{-10pt}
\end{figure}

{\figurename} \ref{figure1} plots the values of $g(\mathbf f)$ as a function of the number of iterations, demonstrating the average convergence performance of the proposed alternating optimization-based method. It is observed that this method converges in approximately $3$ iterations, indicating a fast convergence rate. Additionally, it is observed that increasing the number of antennas $N$ decreases $g(\mathbf f)$. This is because, as the number of antennas increases, the channel correlation between Bob and Eve gradually decreases.

\begin{figure}[!t]
    \centering
    \subfigbottomskip=0pt
	\subfigcapskip=-5pt
\setlength{\abovecaptionskip}{0pt}
   \subfigure[Instantaneous (${R}=10$ bps/Hz).]
    {
        \includegraphics[height=0.175\textwidth]{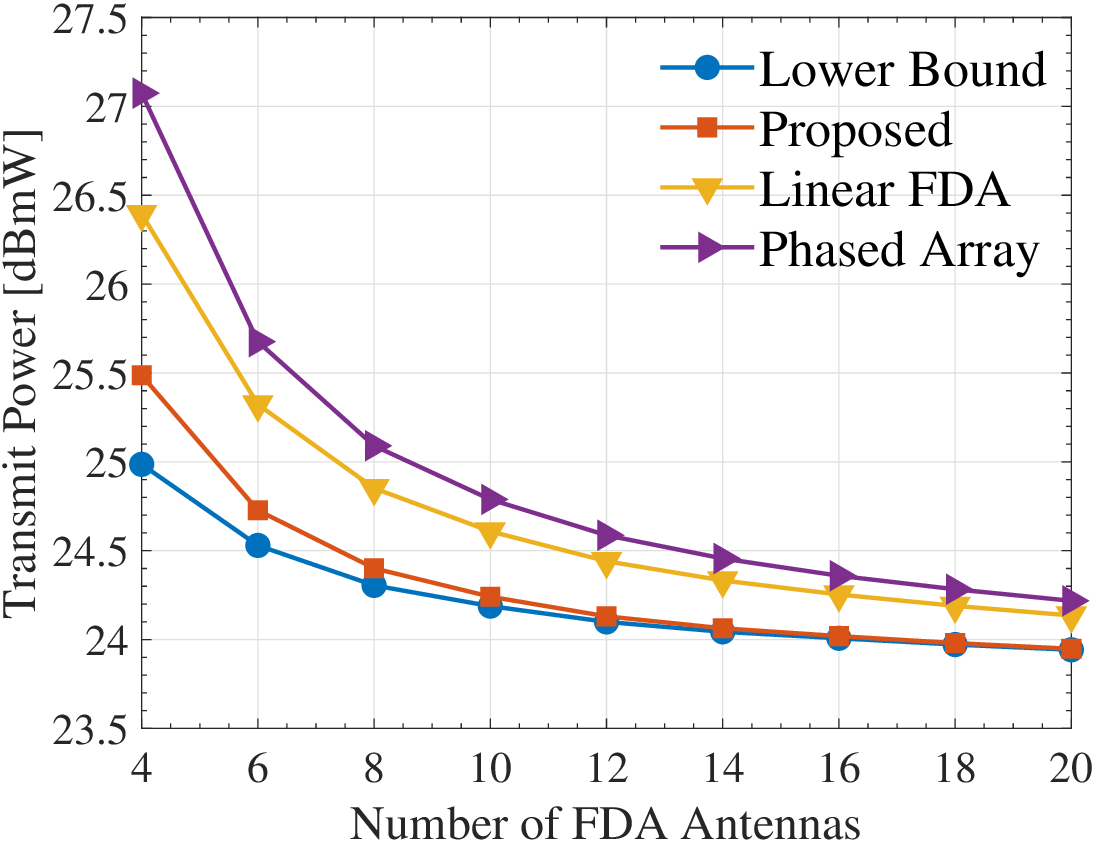}
	   \label{fig2a}	
    }
    \subfigure[Average.]
    {
        \includegraphics[height=0.175\textwidth]{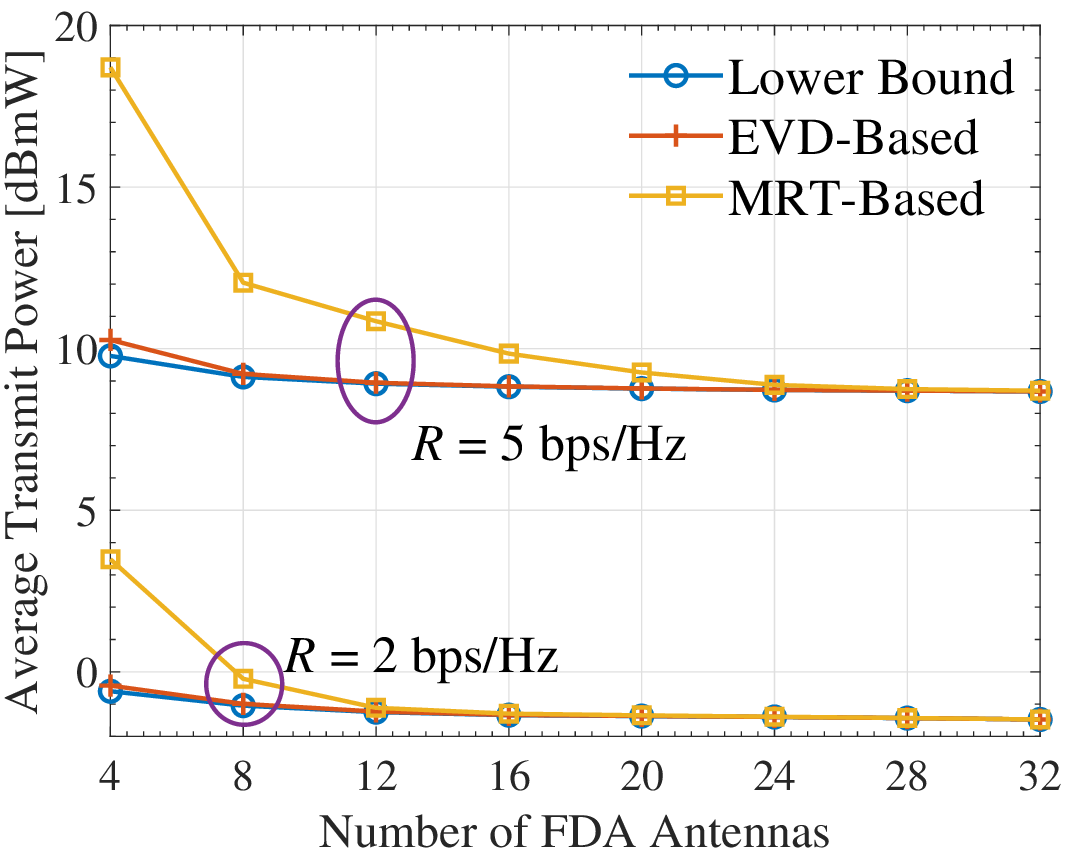}
	   \label{fig2b}	
    }
   \caption{Transmit power minimization.}
    \label{figure2}
    \vspace{-10pt}
\end{figure}

\begin{figure}[!t]
    \centering
    \subfigbottomskip=0pt
	\subfigcapskip=-5pt
\setlength{\abovecaptionskip}{0pt}
   \subfigure[Instantaneous ($N=3$).]
    {
        \includegraphics[height=0.175\textwidth]{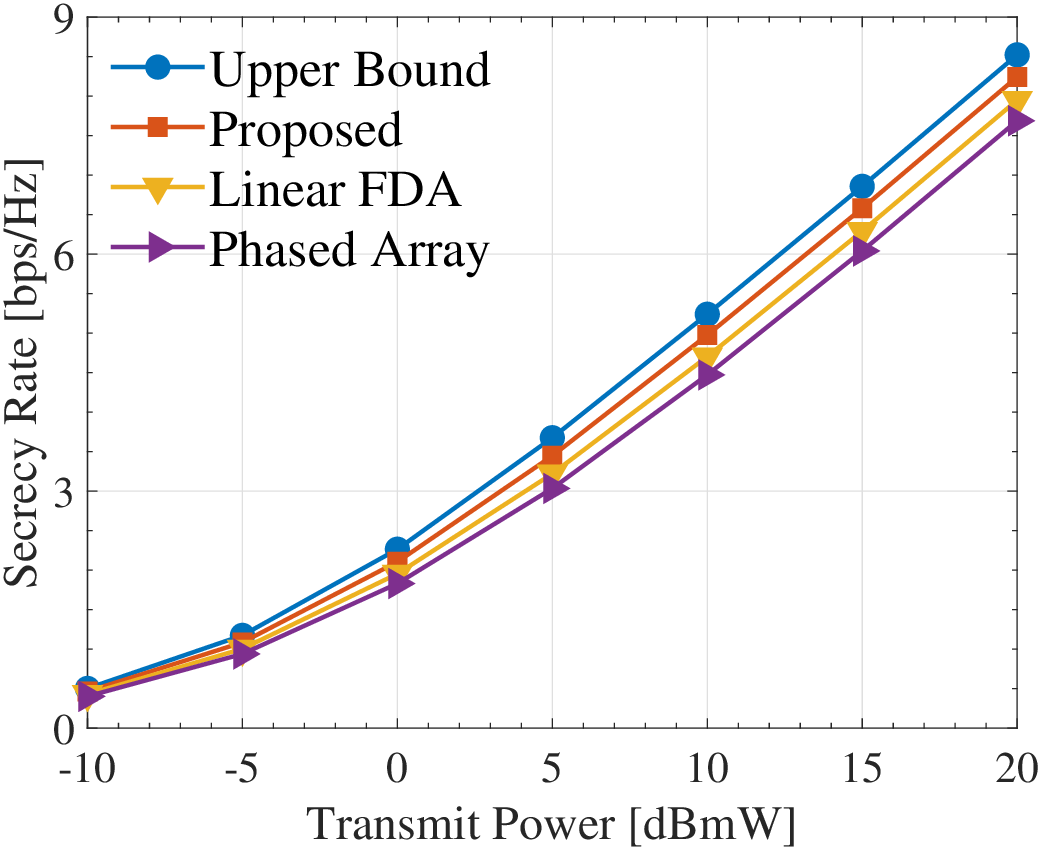}
	   \label{fig3a}	
    }
    \subfigure[Average.]
    {
        \includegraphics[height=0.175\textwidth]{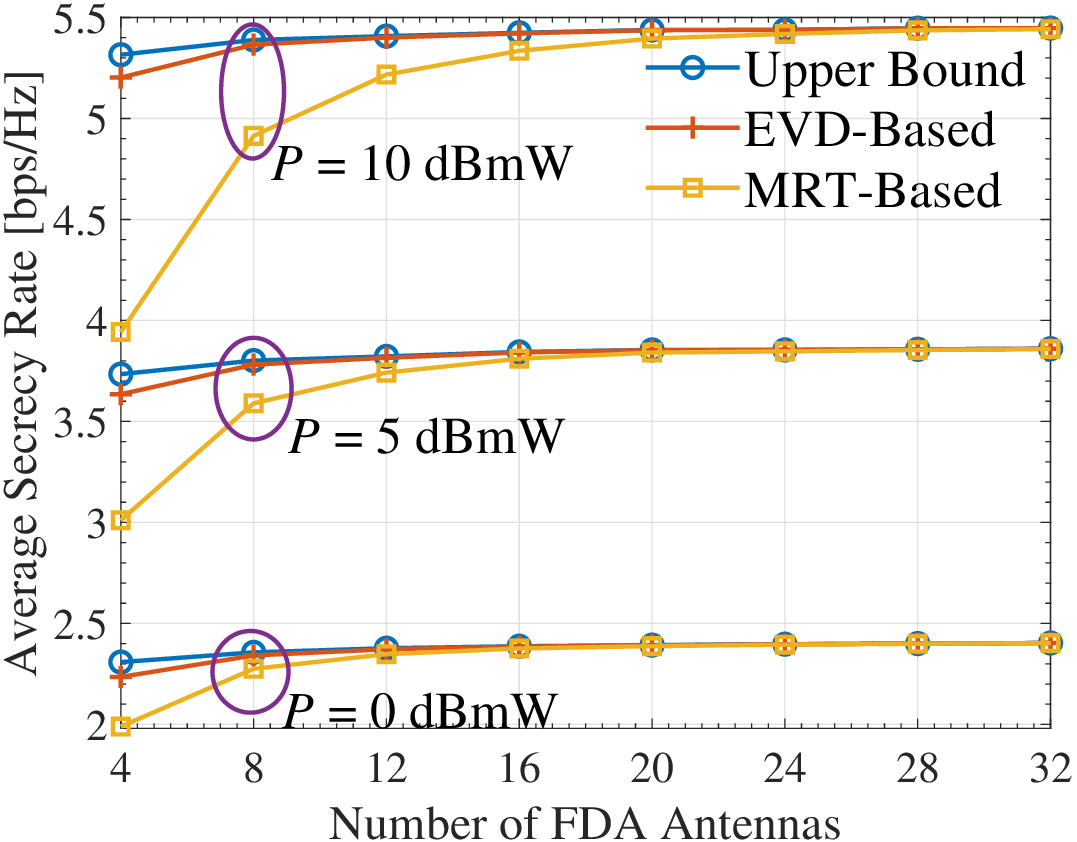}
	   \label{fig3b}	
    }
   \caption{Secrecy rate maximization.}
    \label{figure3}
    \vspace{-10pt}
\end{figure}

{\figurename} \ref{fig2a} illustrates the required transmit power achieved by our proposed method versus the number of FDA antennas. For comparison, we also plot the results obtained by the FDA with linearly varying frequency offsets ($\Delta f_n=\frac{n}{N}f_{\mathsf{m}}$), the phased array ($\Delta f_n=0$), and the performance lower bound ($g(\mathbf{f})=0$). It can be observed that our proposed method outperforms both the linear FDA and phased array schemes. Additionally, as the number of antennas increases, our proposed method converges quickly to the performance lower bound. In {\figurename} \ref{fig2b}, we present the results for the average transmit power minimization problem. It can be seen that using the EVD-based beamforming scheme brings the consumed power close to its lower bound for most values of $N$. The proposed MRT-based scheme also effectively reduces the transmit power, and its performance gradually approaches that of the EVD-based method as $N$ increases. This is because, as $N$ increases, $g(\mathbf{f})$ decreases, and the two schemes eventually achieve similar performance. Next, we turn to the secrecy rate. {\figurename} \ref{fig3a} plots the secrecy rates achieved under different schemes versus the transmit power. As shown, the secrecy rate increases with the available power, and our proposed method is the closest to the performance upper bound ($g(\mathbf{f})=0$). Besides, the results of the average secrecy rate maximization are shown in {\figurename} \ref{fig3b}. As illustrated, our proposed method effectively approaches the upper bound of the secrecy rate, similar to the results in {\figurename} \ref{fig2b}. These results underscore the effectiveness of our FDA-based beamforming scheme in enhancing wireless security.
\section{Conclusion}
We have proposed an FDA-based transmission framework to enhance secrecy transmission. Under the criteria of transmit power minimization and secrecy rate maximization, we introduced a novel, low-complexity method to optimize the FDA frequency offsets. Numerical results validated the effectiveness of the proposed algorithms and demonstrated the superiority of the FDA-based framework over existing techniques.

\end{document}